\documentclass[linenumbers]{aastex631}

\usepackage{url} 
\usepackage{lineno}
\usepackage{bm}

\nolinenumbers

\newcommand{\thbn}{\theta_{\rm Bn}}

\newcommand{\ompe}{\omega_\mathrm{pe}}
\newcommand{\ompi}{\omega_\mathrm{pi}}
\newcommand{\omce}{\Omega_\mathrm{e}}
\newcommand{\omci}{\Omega_\mathrm{i}}

\newcommand{\ms}{M_\mathrm{s}}

\newcommand{\ma}{M_\mathrm{A}}
\newcommand{\mms}{M_\mathrm{fast}}
\newcommand{\mi}{m_\mathrm{i}}
\newcommand{\me}{m_\mathrm{e}}
\newcommand{\lse}{\lambda_\mathrm{se}}
\newcommand{\lde}{\lambda_\mathrm{De}}  
\newcommand{\lsi}{\lambda_\mathrm{si}}
\newcommand{\vsh}{v_\mathrm{sh}}
\newcommand{\vdr}{v_\mathrm{dr}}
\newcommand{\vth}{v_\mathrm{th}}

\newcommand{\nppc}{N_\mathrm{ppc}}

\def\be{\begin{equation}}
\def\ee{\end{equation}}
\def\l{\left}
\def\r{\right}

\def\apjl{ApJ Lett.}%
\usepackage{color}
\usepackage[normalem]{ulem}  

\usepackage{amsmath}  

\newcommand{\art}{\textcolor{black}}
\newcommand{\aar}{\textcolor{black}}

\newcommand{\artrm}{\textcolor{black}}

\newcommand{\rev}{\textcolor{black}}

\newcommand*{\abt}{\mathord{\sim}} 
\newcommand*{\mt}{\mathrm}
\newcommand*{\prll}{\parallel}
\newcommand*{\unit}[1]{\;\mt{#1}}  

\newcommand*{\bp}{\beta_\mathrm{p}}
\newcommand*{\thetaBn}{\theta_{Bn}}  

\newcommand*{\dtl}{\mathrm{d}}

\renewcommand{\vec}[1]{\bm{#1}}  
\newcommand*{\uvec}[1]{\hat{\bm{#1}}}  

\begin{document}

\title{Electrostatic Waves and Electron Holes in Simulations of Low-Mach Quasi-Perpendicular Shocks}

\author[0000-0002-0786-7307]{Artem Bohdan}
\affiliation{Max-Planck-Institut für Plasmaphysik, Boltzmannstr. 2, DE-85748 Garching, Germany}
\affiliation{Excellence Cluster ORIGINS, Boltzmannstr. 2, DE-85748 Garching, Germany}

\author[0000-0003-3483-4890]{Aaron Tran}
\affiliation{University of Wisconsin--Madison Department of Physics, 1150 University Ave, Madison, WI 53706, USA}

\author[0000-0002-1227-2754]{Lorenzo Sironi}
\affiliation{Columbia University Department of Astronomy, 538 W 120th St.~MC~5246, New York, NY 10027, USA}

\author[0000-0002-4313-1970]{Lynn B. Wilson III}
\affiliation{NASA Goddard Space Flight Center, Heliophysics Science Division, Greenbelt, MD, USA}

\begin{abstract}
Collisionless low Mach number shocks are abundant in astrophysical and space plasma environments, exhibiting complex wave activity and wave-particle interactions. In this paper, we present 2D Particle-in-Cell (PIC)  simulations of quasi-perpendicular nonrelativistic ($\vsh \approx (5500-22000)$ km/s) low Mach number shocks, with a specific focus on studying electrostatic waves in the shock ramp and the precursor regions. In these shocks, an ion-scale oblique whistler wave creates \rev{a configuration with} two hot counter-streaming electron beams, which drive unstable electron acoustic waves (EAWs) that can turn into electrostatic solitary waves (ESWs) at the late stage of their evolution. By conducting simulations with periodic boundaries, we show that EAW properties agree with linear dispersion analysis. The characteristics of ESWs in shock simulations, including their wavelength and amplitude, depend on the shock velocity. When extrapolated to shocks with realistic velocities ($\vsh \approx 300$ km/s), the ESW wavelength is reduced to one tenth of the electron skin depth and the ESW amplitude is anticipated to surpass that of the quasi-static electric field by \rev{more than a factor of 100}. These theoretical predictions may explain a discrepancy, between PIC and satellite measurements, in the relative amplitude of high- and low-frequency electric field fluctuations. 
\end{abstract}


\section{Introduction} \label{sec:intro}

Shock waves are ubiquitous both in astrophysical environments and the Solar System. They convert the bulk kinetic energy of supersonic plasma flows into the thermal energy of plasma and facilitate the production of high energy particles, also known as cosmic rays. In most cases, the plasma involved can be treated as collisionless, therefore the energy exchange between plasma species inside the shock transition is governed by collective plasma behaviour and wave-particle interaction. Earth's bow shock provides an excellent laboratory for studying these aspects of shock physics.  Over the past 60 years, it has been extensively investigated in-situ by various satellite missions, such as Cluster \citep{Horbury2001} and Magnetospheric Multiscale Mission \cite[MMS]{Burch2016}. These missions aim to study the microphysics of the Earth's magnetosphere, including the behavior of individual particles and fields at small scales, which is crucial for understanding fundamental processes such as magnetic reconnection, plasma turbulence, particle acceleration, etc.

The most recent mission, MMS, has made approximately 3000 passes through Earth's bow shock \citep{Lalti22}. MMS has provided detailed measurements of electromagnetic fields, wave activity, plasma density, and high-energy particle distributions in the vicinity of the shock.
However, satellite in-situ measurements are limited to the spacecraft's trajectory, providing only a partial description of the shock's three-dimensional structure. As a result, combining these measurements with kinetic plasma simulations can significantly enhance our understanding. Fully kinetic methods, such as Particle-in-Cell (PIC) simulations, have the capability to describe the evolution of shocks at ion scales and resolve the dynamics of electrons. Nevertheless, some discrepancies persist between kinetic simulations and in-situ measurements. In this paper, we want to address the issue raised recently in \cite{Wilson21}, namely, why in real shocks, small-scale electrostatic fluctuations have much larger amplitude than quasi-static electric fields, in contrast to the findings of PIC simulations.

Electrostatic waves of different kinds are detected in-situ near collisionless shocks. They include lower hybrid waves \citep{Tidman71,Wu84,Papadopoulos85,Walker08}, ion acoustic waves (IAWs) \citep{Fredricks68,Fredricks70,Gurnett77,Kurth79,chen2018,davis2021,vasko2022}, electrostatic solitary waves (ESWs) of both positive and negative polarity \citep{bale1998,Behlke04,Wilson07,Wilson10,Wilson14b,goodrich2018,malaspina2020,wang2021}, waves radiated by the electron cyclotron drift instability \citep{Forslund70,Lampe72,Wilson10}, and Langmuir waves \citep{Gurnett77,Filbert79,goodrich2018}. For more detail, see \cite{Wilson21} and citations therein.
Some of them, e.g., IAW and ESW, are characterised by high frequencies and very short wavelengths \citep{Wang21,Vasko22}. \art{Their typical amplitude $E \approx 100$--$200\unit{mV/m}$ is about $50$--$100$ times higher than a typical convective (aka motional) electric field $\lesssim 4 \unit{mV/m}$ measured in the satellite frame \cite[Figure~1]{Wilson21}.
}
Their wavelength is just a few tens of the Debye length or even smaller ($\lambda < 20 \lde \approx 0.1 \lse \approx 170m$, where $\lse$ is the electron skin depth and $\lde$ is the electron Debye length).

In PIC simulations we also can find a number of electrostatic instabilities. For instance, IAWs can be driven by the drift motion of preheated incoming ions relative to the decelerated electrons at the shock foot of high Mach number perpendicular shocks \citep{Kato10a,Kato10b}.
Depending on the shock configuration, EAWs can be observed both in the shock foot as a result of the modified two-stream instability \citep{Matsukiyo06}, or in the \artrm{upstream region} of oblique shocks \artrm{where they are} excited by high-energy electrons moving back upstream \citep{Bohdan22,Morris22}.
Another example involves the excitation of electrostatic waves on the electron Bernstein mode branch by an ion beam \citep{Dieckmann00} when electron cyclotron drift instability becomes dominant. This excitation results in electrostatic waves at multiple electron cyclotron harmonic frequencies \citep{Muschietti06,Yu22} within moderate Mach number perpendicular shocks.
Furthermore, electrostatic Langmuir waves can be generated through the electron bump-on-tail instability \citep{Sarkar15} in the \artrm{upstream} region of oblique high-beta shocks \citep{Kobzar21}.
Buneman instability \citep{Buneman58} occurs between shock-reflected ions and cold upstream electrons, primarily at the shock foot of quasi-perpendicular high \citep{Shimada00,Hoshino02,amano2007,Amano09,Bohdan17,Bohdan19a,Bohdan19b} and low \citep{Umeda09} Mach number shocks.  In most of \artrm{these} PIC simulations, the wavelength of electrostatic waves is comparable to the electron skin depth $\lambda \approx (1-5) \lse \gg \lde$ of the upstream plasma. Additionally, the amplitude of these waves at maximum is typically only a few times larger than \art{the upstream motional electric field ($E_0$) appearing in simulations performed in the downstream rest reference frame, $E/E_0 \lesssim 2$}, which appears inconsistent with satellite measurements. \aar{Note that, for non-relativistic shock simulations performed
    in the downstream plasma's rest frame,
    $E_0$ differs from the normal incidence frame (NIF)
    motional electric field by a factor of 
    $(1 - 1/r)$,
    where $r$ is the density compression ratio.
    Following \cite{Wilson21}, we treat $E_0$ as 
    comparable to spacecraft-frame measurements
    of the solar wind motional field
    within a factor of a few.}

A potential explanation for the observed discrepancies between simulation results and in-situ measurements lies in the choice of simulation parameters. In many cases, simulations adopt parameters that are unrealistic in order to ensure computational feasibility. Therefore, it is crucial to understand how the physical picture is distorted within simulations to accurately describe real systems. Depending on the problem in question, a range of correction techniques may be required for meaningful comparisons with in-situ measurements. These can vary from minimal corrections, when magnetic field amplification by Weibel instability is considered \citep{Bohdan21}, to more intricate rescaling calculations for problems of electron heating \citep{Bohdan20b} or kinetic plasma waves \citep{Verscharen20}, particularly when unrealistically high shock velocities or low ion-to-electron mass ratios are employed.
In shock simulations, electrostatic waves can arise from various two-stream instabilities between drifting plasma components (ion-ion, ion-electron, electron-electron). In such cases, the parameters of these waves could depend on the \art{relative} drift velocity \aar{between} \art{plasma components}. Since the \art{energy} source of the \art{relative} plasma drift is the upstream plasma's \aar{bulk flow kinetic energy}, the drift velocity \art{could} be roughly proportional to the shock velocity. Therefore, if a realistic shock velocity is utilized in a simulation, electrostatic waves
may have
different wavelength and amplitude \art{than for typical PIC simulation parameters (unrealistically high shock velocities and low ion-to-electron mass ratios)}. Here,
we aim to test this idea using PIC simulations and linear dispersion analysis.

The paper is structured as follows. Section 2 is dedicated to shock simulations. In Section 3 we discuss the results of the linear dispersion analysis and PIC simulations with periodic boundaries representing local regions within a shock. In Section 4 we discuss our results, and Section 5 summarizes our findings.

\section{Shock simulations}

\subsection{Simulation setup} \label{pic_setup}

We use the particle-in-cell (PIC) code TRISTAN-MP
\citep{buneman1993,spitkovsky2005} to
simulate 2D quasi-perpendicular shocks with
sonic Mach number $\ms = 4.0$,
Alfv\'{e}n Mach number $\ma = 1.8$,
fast-mode Mach number \art{$\mms=1.68$},
upstream total plasma beta $\bp = 0.25$,
and upstream magnetic field angle $\theta_\mathrm{Bn}=65^{\circ}$ with respect to the
shock-normal coordinate.
The upstream magnetic field lies within the simulation
plane.
The same shock parameters ($\mms$, $\bp$, $\thetaBn$) were studied by \cite[Section~7]{tran2023-oblique}; here, we branch off of their work
using targeted 2D simulations \aar{to study electrostatic wave properties}.

We form a shock by initializing a thermal plasma with bulk velocity
$\vec{v}_0 = - v_0 \uvec{x}$, single-species density $n_0$, and upstream temperature $T_0$.
The plasma has two species: ions and electrons.
The moving upstream plasma carries magnetic field
$\vec{B}_0$ and electric field
$\bm{E}_{\rm 0}=-\bm{v}_{\rm 0}\times\bm{B}_{\rm 0}$,
where $\bm{v}_{\rm 0}$ is
the upstream plasma velocity in the simulation reference frame.
The \artrm{upstream} plasma reflects on a conducting wall
at $x=0$,
and the reflected plasma interacts with upstream plasma
to form a shock traveling
towards $+\uvec{x}$.
The simulation proceeds in
approximately the downstream (i.e., post-shock) plasma's rest frame, except for
a \artrm{small} drift in the shock-transverse direction that is expected for oblique
shocks \citep{Tidman71}.
\aar{The far $x$ boundary continuously expands towards $+\uvec{x}$ and injects fresh plasma into the simulation domain \citep{sironi2009}.
The domain's $y$ boundaries are periodic.}

The shock speed $\vsh$---i.e., the upstream flow speed in the shock's rest frame---is not directly chosen.
We compute it as $\vsh = v_0 / (1-1/r)$ in the non-relativistic limit,
with $r$ the density compression ratio estimated from
the oblique magnetohydrodynamic (MHD) Rankine-Hugoniot conditions
assuming adiabatic index $\Gamma=5/3$ \citep{Tidman71}.
By numerically inverting this procedure, we can choose $v_0$ to target a desired
$\vsh$ and hence Mach number.
\rev{
    To relate $\vsh$ and $v_0$ in Table~\ref{tab_shock_param},
    we use $r=1.8496$.
}
The targeted and actual Mach numbers agree to within $\sim2$--$7\%$.
A more detailed explanation is given in \cite{tran2023-oblique}.

Standard plasma lengthscales and timescales are defined using upstream
(pre-shock) plasma quantities; we use SI (MKS) units.
The electron plasma frequency $\ompe = \sqrt{n_0 e^2/(\epsilon_0 \me )}$,
the electron cyclotron frequency $\omce = e B_0 /\me $,
the electron skin depth $\lse = c/\ompe$,
and the electron Debye length $\lde = \sqrt{\epsilon_0 k_B T_0/(n_0 e^2)}$.
Here,
$e$ is the elementary charge and $\epsilon_0$ is the vacuum permittivity.
Ion quantities $\ompi$, $\lsi$, $\omci$ are defined analogously.
We define \aar{Mach numbers}
$\ms = \vsh/c_\mt{s}$, $\ma = \vsh/v_\mt{A}$,
and \art{$\mms=\vsh/v_\mt{fast}$}.
\aar{
    The shock speed $\vsh$ is the upstream flow speed
    measured in the shock's rest frame, the
}
sound speed $c_\mt{s}=\sqrt{2\Gamma k_\mt{B} T_0/(\mi+\me)}$,
the Alfv\'{e}n speed $v_\mt{A}=B_0/\sqrt{\mu_0 n_0 (\mi+\me)}$,
and \art{
the MHD fast speed
$v_\mt{fast}=\sqrt{0.5(c_\mt{s}^2+v_\mt{A}^2+\sqrt{(c_\mt{s}^2+v_\mt{A}^2)^2-4c_\mt{s}^2 v_\mt{A}^2 \cos^2{\theta_\mathrm{Bn}}})}$.
}
The total plasma beta $\bp = 4 \mu_0 n_0 k_\mt{B} T_0/B_0^2$.
The constants $\mi$ and $\me$ are ion and electron masses,
$k_\mt{B}$ is the Boltzmann constant,
$c$ is the speed of light,
and $\mu_0$ is the vacuum magnetic permeability.
We define the initial electron root-mean-square thermal velocity
$v_\mt{te0} = \sqrt{k_\mt{B} T_\mt{0}/\me}$.

Fixing the shock parameters ($\mms$, $\bp$, $\thetaBn$), we vary
$v_\mt{sh}/c$ and the ion-electron mass ratio $\mi/\me$
to study how the resulting shock structure depends upon numerical compromises
adopted for PIC simulations
(Table~\ref{tab_shock_param}).
\aar{
To vary $\vsh/c$, we rescale the dimensionless parameters
$v_0/c$, $k_\mathrm{B} T_0/(m_i c^2)$, and $v_\mathrm{A}/c$,
which are used to inject upstream plasma.
If the flow and thermal speeds are non-relativistic,
we anticipate that the shock's macroscopic behavior may not depend
on $\vsh/c$, or any other quantity scaled with respect to $c$ (including $\ompe/\omce$ and $\ompi/\omci$), so long as the dimensionless parameters $\ms$, $\bp$, $\thetaBn$, and $\mi/\me$ are fixed.
Thus, PIC simulations with large $\vsh \sim 10^4 \unit{km/s}$
(and hence large $T_0$) may serve as analogs
for natural systems with lower flow speeds $\sim 10^2 \unit{km/s}$.
In our simulations, all speeds are non-relativistic except for the electron thermal speed, which can be $\sim 0.1c$ (but the electron thermal energy remains $\ll \me c^2$).
However, electron-scale waves may be sensitive to $\vsh/c$ (equivalently, $\ompe/\omce$); these waves could in principle have a global effect upon shock structure.
It is this subtler dependence that we seek to study.
}

All simulations have transverse width $L_y = 38.4\lse$ ($0.9$ to $2.9 \lsi$),
duration $8.5 \omci^{-1}$, and
spatial grid resolution $\Delta = 1.0 \lde$.
The upstream plasma temperature $k_B T_0 = 10^{-4}m_\mathrm{i}c^2$
for Run A and scales with $(\vsh/c)^2$ for other runs so as to fix $\ms=4$.
The electron plasma-to-cyclotron frequency ratio $\ompe/\omce = 1.8$--$7.0$ for
Runs A--E respectively. The runs of varying mass ratio (B, B400, B800, B1836)
all have $\ompe/\omce = 2.49$. Note that for fixed $\bp$,
$\ompe/\omce$ scales linearly with $v_\mt{sh}/c$,
and $\lde/\lse$ scales inversely with $v_\mt{sh}/c$.

The runs in Table~\ref{tab_shock_param} use
$128$ particles per cell ($64$ per species), but we also perform variant simulations with up to $512$ per cell ($256$ per species) to test convergence.
We smooth the PIC current with 32 passes of a digital ``1--2--1'' filter on
each coordinate axis, which imposes 50\% power damping at wavenumber
$k_\mt{damp} \approx 0.25 \left( \Delta \right)^{-1}$ \citep[Appendix~C]{birdsall1991}.

\begin{table}[]
    \centering
    \begin{tabular}{lcccccc}
    \hline
        Run     & $m_\mt{i}/m_\mt{e}$ & $\vsh/c$ & $v_0/c$
        & Width ($\lsi$) & $\lde/\lse=v_\mt{te0}/c$ & $\ompe/\omce$ \\
    \hline
A       & 200   & 0.0733 & 0.0338 & 2.90 & 0.143 & 1.76 \\
B       & 200   & 0.0518 & 0.0238 & 2.71 & 0.100 & 2.49 \\
C       & 200   & 0.0366 & 0.0168 & 2.90 & 0.071 & 3.52 \\
D       & 200   & 0.0259 & 0.0119 & 2.71 & 0.050 & 4.99 \\
E       & 200   & 0.0183 & 0.0084 & 2.90 & 0.036 & 7.05 \\
    \hline
B400    & 400   & 0.0367 & 0.0169 & 1.92 & 0.100 & 2.49 \\
B800    & 800   & 0.0260 & 0.0119 & 1.36 & 0.100 & 2.49 \\
B1836   & 1836  & 0.0171 & 0.0079 & 0.90 & 0.100 & 2.49 \\
    \hline
    \end{tabular}
    \caption{
        Shock simulation parameters.
        All simulations have $\ms=4$, $\ma=1.8$, and $\thbn=65^\circ$.
        Columns are the ion-to-electron mass ratio $\mi/\me$, the shock velocity $\vsh/c$, the upstream velocity in the lab-rest frame $v_0/c$, the simulation domain width $L_y$ measured in terms of the ion skin depth $\lsi$,
        the ratio of upstream electron Debye and skin lengths $\lde/\lse$ (which equals $v_\mt{te0}/c$),
        and the electron plasma-to-cyclotron frequency ratio $\ompe/\omce$.
        \rev{
        To relate $\vsh$ and $v_0$,
        we use $r=1.8496$.
        }
    }
    \label{tab_shock_param}
\end{table}

\subsection{Shock properties}

\begin{figure}
\centering
\includegraphics[width=\linewidth]{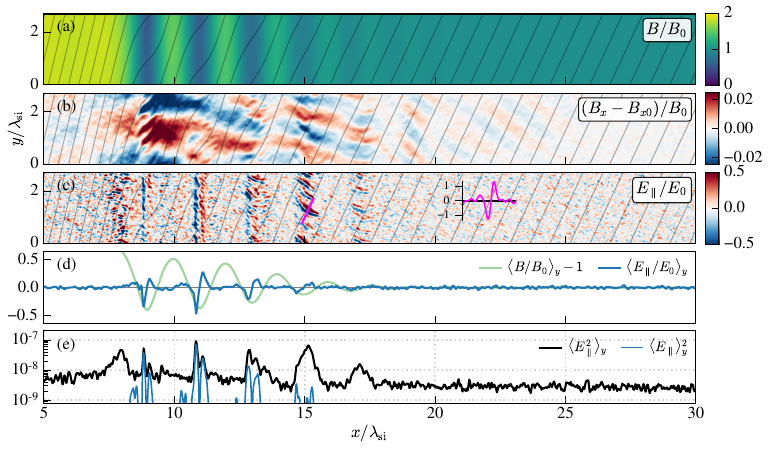}
\caption{
    Overview of Run B at $t=8.50{\omci}^{-1}$.
    The shock travels from left to right
    and has its shock ramp
    at $x \approx 8\lsi$, \art{which coincides with the largest density jump}.
    (a) Magnetic field magnitude  \aar{$B/B_0$}.
    (b) Shock-normal magnetic fluctuation $(B_x-B_{x0})/B_0$, scaled to
        upstream field values $B_0$ and $B_{x0}$.
    (c) Parallel electric field, $E_\prll/E_0$, measured with respect to local $\vec{B}$
    \aar{and scaled to upstream motional electric field $E_0$.}
    \aar{
    Inset waveform plot (magenta curve) shows $E_\parallel/E_0$ waveform of a strong electron hole,
    measured along the magenta ray at $x = 15\lsi$.
    Abscissa increases along $\vec{B}$ and towards $+\uvec{x}$.
    (d) The $y$-averaged parallel electric field
    $\langle E_\parallel/E_0\rangle_y$ (blue)
    and magnetic fluctuation
    $\langle B/B_0 \rangle_y -1$ (green).
    (e) Total (black) and $y$-averaged (blue) parallel electric field energy densities in arbitrary units.
    }
    In panels (a)-(c), black contours trace magnetic field lines.
}
\label{overview}
\end{figure}

Figure~\ref{overview} shows the structure of Run B at $t = 8.50{\omci}^{-1}$,
which exemplifies the general structure of all the simulations in
Table~\ref{tab_shock_param}.
The shock speed $v_\mt{sh}$ is less than the phase speed of oblique
whistlers traveling along shock normal ($+\uvec{x}$), allowing a
phase-standing precursor wave train to form ahead of the shock \citep{krasnoselskikh2002}.

In Figure~\ref{overview}(a), the main $\vec{B}$-field compression (i.e., the
shock ramp) takes place at $x\sim 8\lsi$, with compressive oscillations at
larger $x$ that gradually decay in the $+\uvec{x}$ direction.
Figure~\ref{overview}(b) shows the shock-normal magnetic fluctuation $(B_x - B_{x0})/B_0$; recall that the upstream field component
$B_{x0} \equiv B_0 \cos\thetaBn$ is conserved across the shock jump.
The $B_x$ fluctuation reveals electromagnetic waves with $\vec{k}$ oriented
oblique to both $\vec{B}_0$ and shock normal
(i.e., not phase-standing), which we call the oblique precursor.
The oblique precursor has $\abt 100\times$ smaller amplitude
than the $\uvec{x}$-aligned compressive waves in Figure~\ref{overview}(a)).
Both precursor wave trains in Figure~\ref{overview}(a) and
(b) are right-hand polarized, consistent with the fast-mode/whistler
branch of the plasma dispersion relation.

The precursor wave trains at $t=8.50 {\omci}^{-1}$ (Figure~\ref{overview}) are not in steady state.
If the simulation proceeds to longer times $t \sim 40 {\omci}^{-1}$, then
(i) the oblique precursor grows in amplitude,
(ii) both the phase-standing and oblique precursors extend farther ahead of the shock, and
(iii) density filamentation appears within and ahead of the shock ramp
\citep{tran2023-oblique}.
We emphasize that our shock simulations are deliberately
shorter in duration (not steady-state) and also narrower in transverse width
than some other fully-kinetic PIC simulations in the recent literature
\citep{xu2020,lezhnin2021,bohdan2022,tran2023-oblique}.
The narrow transverse domain width $\abt 3\lsi$ helps preclude or slow
the growth of other ion- and fluid-scale waves that would appear at
shock-transverse scales of \artrm{$\abt 10 \lsi$}
\citep{lowe2003,burgess2016,johlander2016,trotta2023}.
The simulation parameters ensure that
(i) the shock is steady on electron timescales,
and (ii) its overall structure is
dominated by a single, coherent precursor wave train without other ion-scale waves interfering.
\rev{It aids our analysis to isolate a single ion-scale wave mode that then forms electrostatic solitary waves; in real shocks, multiple ion-scale waves may exist in superposition to dictate the shock's behavior.}

Figure~\ref{overview}(c) shows 
\aar{$E_\prll \sim E_0$} fluctuations co-existing
with the \art{whistler} precursor waves,
\aar{where $E_0 = v_0 B_0$ is the magnitude of the upstream motional electric field.}
A laminar component
\aar{with}
$\vec{k} \parallel +\uvec{x}$
\aar{appears at $x \sim 9$ to $13\lsi$},
and smaller bipolar
structures with \aar{$\vec{k} \parallel \vec{B}$}
\aar{prevail at $x \sim 13$ to $18\lsi$}.
The bipolar structures have positive polarity: $E_\parallel$ points away
from the center of the structure, so the electric potential
$\phi(s) = - \int_{-\infty}^{s} E_\parallel(s') \dtl s'$
has local maximum at the structure's center,
with $s$ a magnetic field-aligned coordinate.
\aar{
The typical 1D $E_\parallel$ profile of such a structure is shown as an inset (magenta curve) in Figure~\ref{overview}(c).}
We refer to these electrostatic fluctuations as 
electron holes \rev{or ESWs (we use both names interchangeably)}, anticipating that they represent the non-linear outcome of an electron-electron streaming instability to be shown in Section~\ref{sec:lin_disp_an}.
\rev{The ``hole'' refers to a void in electron velocity space that forms within the self-consistent bipolar electrostatic fields \citep{hutchinson2017}.}

\aar{
Figure~\ref{overview}(d) shows the 1D $y$-averaged profiles $\langle E_\parallel \rangle_y$ and magnetic fluctuation $\langle B/B_0 \rangle_y - 1$;
we denote $y$-averaging by
$\langle \cdots \rangle_y$.
The $E_\parallel$ fluctuations near the shock
with $\vec{k} \parallel +\uvec{x}$ form positive electrostatic potentials within
\rev{the low-$B$ parts of the precursor wave's cycle, which we call}
magnetic troughs.
Figure~\ref{overview}(e) shows the total electrostatic parallel energy density $\langle E_\parallel^2\rangle_y$ (black).
At $x \sim 9$ to $13\lsi$, we see that
$\langle E_\parallel^2\rangle_y$
mostly arises from the $y$-averaged energy density
$\langle E_\parallel\rangle_y^2$ (blue),
which captures $E_\parallel$ fluctuations with $\vec{k} \parallel +\uvec{x}$.
Left of the shock ramp, and right of $x \sim 13 \lsi$, we see that
$\langle E_\parallel^2 \rangle_y$ arises from short-wavelength fluctuations not captured in $\langle E_\prll \rangle_y^2$.
}

\begin{figure}
    \centering
    \includegraphics[width=.9\linewidth]{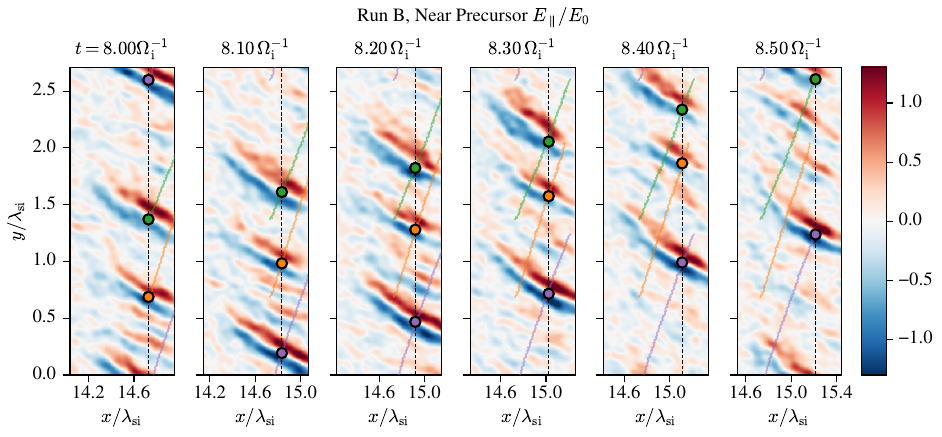}
    \caption{
        Three electron hole trajectories in Run B, tracked between $8.0$ and
        $8.5 \,\omci^{-1}$.
        The $E_\parallel/E_0$ panels advance in time from left to right,
        co-moving with the shock's precursor wave train; the $x$ and $y$ axis
        coordinates are measured in the simulation frame.
        Individual electron holes are marked by green, orange, and purple dots.
        The faint lines of corresponding color show their trajectories over time.
        The ``Near Precursor'' region shown is defined in
        Section~\ref{sec:powerscaling}.
    }
    \label{ehole}
\end{figure}

To show how the electron holes evolve in time,
we track the real-space trajectory of three
example holes in Run B (Figure~\ref{ehole}).
To do so,
we select the magnetic trough at $x\approx14$ to $15\lsi$
and measure a 1D $E_\parallel(y)$ profile
\aar{
    at an $x$ position offset $+\lambda/8$
    from the magnetic trough's minimum
    (Figure~\ref{ehole}, dashed black line),
    where $\lambda$ is the \emph{local}
    ion-scale precursor wavelength.
}
Holes are identified 
\aar{as locations where $E_\prll(y)=0$ in}
between adjacent extrema $|E_\prll|/E_0 > 0.33$.
We track \aar{three manually-chosen} holes from
$t=8.0$ to $8.5\,\omci^{-1}$.
\aar{In the upstream plasma's rest frame,
the hole} velocities are $0.98 v_\mt{te0}$ (orange dot),
$0.87 v_\mt{te0}$ (green dot), and $0.91 v_\mt{te0}$ (purple dot),
\aar{
    recalling that $v_\mt{te0}$ is an upstream electron thermal velocity.
    By construction, all holes have upstream-frame
    $v_x \approx v_\mt{sh}$ within a few percent.
    The upstream-frame $v_y = 3.0 v_\mt{A}$,
    $2.5 v_\mt{A}$, and $2.7 v_\mt{A}$ respectively;
    the velocity vectors have corresponding
    angles $59^\circ$, $54^\circ$, and $56^\circ$ slightly below the local magnetic field angle of $62$--$63^\circ$.
    The upstream-frame velocity $v$ is somewhat less than the Landau resonance velocity $\omega/k_\parallel \approx v_\mt{sh}/\cos\thetaBn = 1.22 v_\mt{te0}$.
}

\aar{
In \ref{app1}, we present data from all of Runs A--E
to establish that electron holes appear with amplitude exceeding PIC noise
and that both $\omega$ and $\vec{k}$ associated with the holes are well separated from other wave modes.
}

\subsection{Electrostatic energy scaling with $v_\mt{sh}/c$ and $\mi/\me$}
\label{sec:powerscaling}

How does the electrostatic energy density and wavenumber vary with $v_\mt{sh}/c$?
To compare these quantities between different simulations, we define four
regions: the ``Ramp'', ``Near Precursor'', ``Far Precursor'', and
``Control'', which are constructed as follows.

\begin{figure}
\centering
\includegraphics[width=\linewidth]{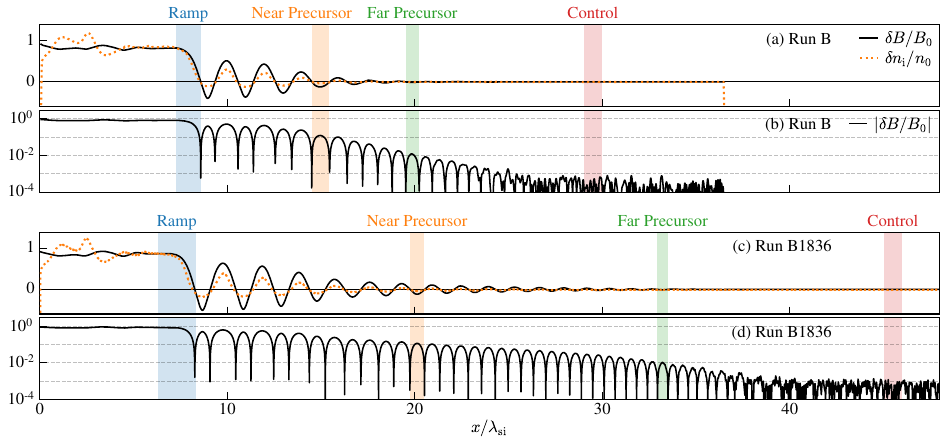}
\caption{
    Electrostatic waves in Runs B and B1836 are measured in the colored
    regions: ``Ramp'' (blue), ''Near Precursor'' (orange), ``Far Precursor''
    (green), and ``Control'' (red).
    In panels (a)--(b), the $y$-averaged
    fluctuations
    \aar{
    $\delta B/B_0$ (black) and
    $\delta n_\mt{i}/n_0$ (orange dotted)
    }
    show the region selection procedure for Run B.
    Panels (c)--(d) illustrate the same procedure for Run B1836.
    The ``Near'' and ``Far Precursor'' regions are magnetic troughs \rev{(wave half-cycles of low amplitude)} with
    $|\delta B/B_0|$ just exceeding $0.1$ and $0.01$ respectively.
    The ``Ramp'' \aar{
        is the contiguous region where
        $\delta n_\mathrm{i}/n_0 < \delta B/B_0$
        (corresponding to $\delta B_\mt{ff} > 0$)
        at the \art{sharpest magnetic field increase} in panels (a) and (c).
    }
    The ``Control'' region samples undisturbed upstream plasma.
}
\label{segment_procedure}
\end{figure}

We segment the 1D, $y$-averaged, magnetic fluctuation strength
$\delta B \equiv \left[B(x)/B_0\right] - 1$
by using its zero crossings to separate the precursor wave into half cycles
\rev{of low and high $B$ amplitude, called troughs and crests respectively}
(Figure~\ref{segment_procedure}(b)).
Within the precursor, ES waves occur in troughs.
The ``Near Precursor'' region is the right-most segment with $\delta B < 0$ and
extremum $|\delta B| \ge 0.1$ within the wave trough.
The ``Far Precursor'' region is the right-most segment with $\delta B < 0$ and
extremum $|\delta B| \ge 0.01$ within the wave trough.
\rev{For the precursor regions,
the dimensionless amplitude thresholds
of $0.1$ and $0.01$ are chosen to select for non-linear versus linear precursor
wave behavior, respectively.}
For the ``Ramp'' region, we
\aar{select an $x$-interval around the sharpest magnetic field increase, wherein magnetic flux-freezing is locally broken such that the magnetic field is compressed more than the density, $\delta n_\mathrm{i}/n_0 < \delta B/B_0$. Let}
$\delta B_\mt{ff} \equiv \left[B(x)/B_\mt{ff}(x)\right] - 1$,
\aar{ defining }
the flux-frozen field $B_\mt{ff}(x) = B_0
n_\mt{i}(x)/n_0$.
We segment $\delta B_\mt{ff}$ using its zero crossings, and we select
the unique region of $\delta B_\mt{ff} > 0$ that coincides with the
\aar{conventionally-defined}
shock ramp;
\aar{ i.e., the largest rise in magnetic field amplitude within the shock transition }
(Figure~\ref{segment_procedure}(a),(c)).
The so-chosen ``Ramp'' region has a similar width
($1.1$ to $1.4 \lsi$) in all of Runs A--E.
For higher mass ratio $\mi/\me$, the precursor wave train
extends for more cycles ahead of the shock.
The region selections are thus spaced farther apart
(Figure~\ref{segment_procedure}\artrm{(c),}(d)).
The ``Control'' region is 
\rev{fixed to the $x$-coordinate}
intervals \rev{29--30}, 35--36, 40--41, and
45--46 $\lsi$ for Runs \rev{A--E}, B400, B800, and B1836 respectively.

\begin{figure}
\centering
\includegraphics[width=.8\linewidth]{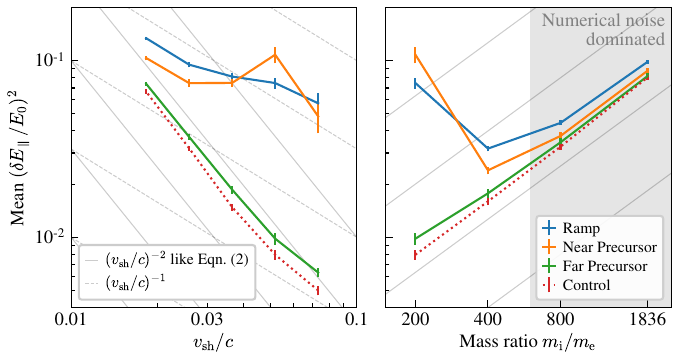}
\caption{
    Left panel: scaling of the mean electrostatic energy density
    $\delta E_\prll^2$ (Equation~\eqref{eprlldens}) with $v_\mt{sh}/c$,
    for four regions in each of Runs A--E.
    Solid gray lines show numerical noise scaling
    \artrm{$(\delta E_\prll/E_0)^2 \propto (\vsh/c)^{-2}$}
    like Equation~\eqref{esw_model_noise};
    dashed gray lines show $(\delta E_\prll/E_0) \propto (\vsh/c)^{-1}$.
    Right panel: scaling with mass ratio $\mi/\me$ based on Runs B, B400, B800 and B1836.
    Solid gray lines show numerical noise scaling
    $(\delta E_\prll/E_0)^2 \propto (\vsh/c)^{-2} \propto \mi/\me$
    appropriate for our simulation parameters.
    The electrostatic energy for Runs B800 and B1836 is dominated by numerical noise.
}
\label{power_scaling}
\end{figure}

\begin{figure}
\centering
\includegraphics[width=\linewidth]{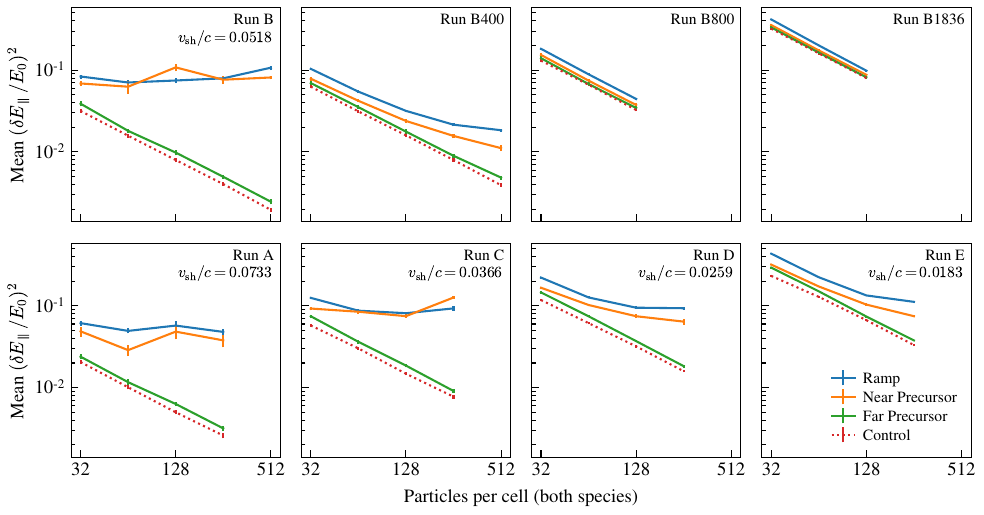}
\caption{
    Like Figure~\ref{power_scaling}, but showing numerical convergence with
    respect to total number of particles per cell, counting both species.
    Runs A, B, and C are converged, based on the near-constant electrostatic
    energy density in the ``Ramp'' and ``Near Precursor'' regions;
    Runs D, E, and B400 are less or not converged.
    Runs B800 and B1836 are not converged and/or do not generate electrostatic
    fluctuations above numerical noise.
}
\label{power_scaling_ppc}
\end{figure}

Figure~\ref{power_scaling} (left panel) shows how the electrostatic energy density
\aar{$(\delta E_\prll/E_0)^2$}
scales with $v_\mt{sh}/c$ in each region.
\rev{
    To improve signal-to-noise, we average $E_\parallel$ over $(x,y,t)$ within
    the time interval
    $t = 8.00$ to $8.50{\omci}^{-1}$.
    And,
}
we measure only fluctuations with wavevector $k_y \ne 0$ by subtracting the
$y$-averaged contribution:
\begin{equation} \label{eprlldens}
    \delta E_\prll^2
    \equiv
    \langle E_\prll^2 \rangle
    - \langle \langle E_\prll \rangle_y^2 \rangle
    =
    \frac{1}{V}
    \iiint E_\prll^2 \,\mathrm{d}y \,\mathrm{d}x \,\mathrm{d}t
    -
    \frac{1}{V}
    \iint \left[\int E_\prll \,\mathrm{d}y \right]^2 \mathrm{d}x \,\mathrm{d}t
    \, ,
\end{equation}
where $V$ is the \aar{3D $(x,y,t)$} integration volume,
\aar{
recalling from Figure~\ref{overview} that the whistler precursor hosts $E_\prll$
fluctuation with both $k_y = 0$ and $k_y \ne 0$.
}
For our chosen regions, $\langle E_\prll^2 \rangle \gg
\langle\langle E_\prll \rangle^2_y \rangle$.
\rev{In Figure~\ref{power_scaling},}
vertical error bars show the standard deviation, in time, of the space-averaged
energy density
in each region.

Both the ``Far Precursor'' and ``Control'' regions show
$(\delta E_\prll/E_0)^2 \propto (v_\mt{sh}/c)^{-2}$
(Figure~\ref{power_scaling}, left panel),
which we attribute to numerical fluctuations.
    All of Runs A--E use the same number of PIC macroparticles per Debye sphere, $\Lambda_\mt{p}$,
    so we expect
$\delta E_\prll^2 \propto n_\mt{e} k_\mt{B} T_\mt{e}$
    up to a constant prefactor that depends on $\Lambda_\mt{p}$ and
    the numerical particle shape \citep[Section~5]{melzani2013}.
Therefore,
\begin{equation} \label{esw_model_noise}
    \left(\frac{\delta E_\prll}{E_0}\right)^2
    \propto \beta_\mt{e} \left( v_\mt{sh}/c \right)^{-2}
    \, ,
\end{equation}
\artrm{
    where $\beta_\mt{e}$ is the electron plasma beta.
}

In contrast, the ``Ramp'' and ``Near Precursor'' regions suggest a different
scaling behavior, $(\delta E_\prll/E_0)^2 \propto (v_\mt{sh}/c)^0$ or
$(v_\mt{sh}/c)^{-1}$.
The scaling may also show a turnover caused by a transition from
mildly-relativistic to non-relativistic regimes, as thermal electrons
attain velocities of $\abt 0.1$ to $0.5c$ in Runs A and B,
\aar{which have $\vsh/c = 0.0733$ and $0.0518$ respectively}.

Mass ratio dependence is shown in the right panel of Figure~\ref{power_scaling}.
The energy density in the ``Ramp'' and ``Near Precursor'' regions decreases
with mass ratio $\mi/\me$ until reaching a numerical noise floor.
For our simulations, the numerical noise
$(\delta E_\prll/E_0)^2 \propto (\vsh/c)^{-2} \propto (\mi/\me)^1$
has an implicit mass ratio scaling
because we hold $v_\mt{te0}/c$ constant while varying mass ratio,
which implies that $\vsh/c \propto (\mi/\me)^{-1/2}$ in Equation~\ref{esw_model_noise}.

Figure~\ref{power_scaling_ppc} checks whether the data in
Figure~\ref{power_scaling} are converged with respect to numerical particle
sampling.
The energy density is mostly converged for $\mi/\me=200$ (Runs A--E), and it is
nearly converged for $\mi/\me=400$ (Run B400).
Higher mass ratios appear dominated by numerical noise.

\begin{figure}
    \centering
    \includegraphics[width=.75\linewidth]{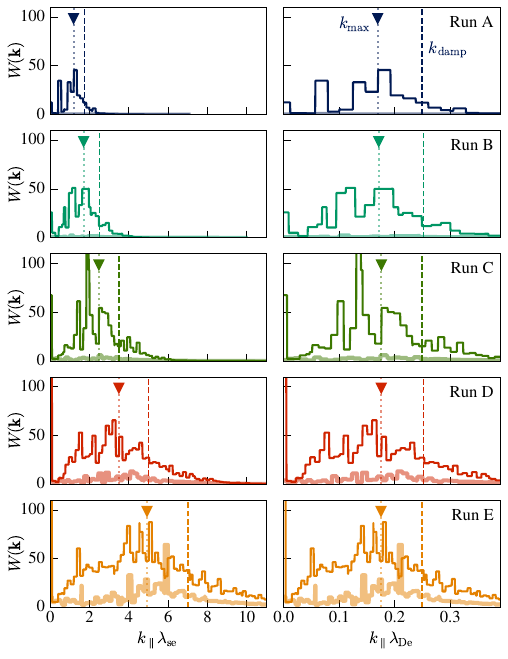}
    \caption{
        Electrostatic power spectrum $W(\vec{k}=k_\parallel\uvec{b})$,
        measured along
        \artrm{the ray $k_\perp = 0$,}
        for the ``Ramp'' region in Runs A--E.
        Left column shows $k_\prll$ scaled to $\lse$,
        right column shows $k_\prll$ scaled to $\lde$.
        Thick translucent lines are ``Control'' region power spectra.
        The triangle and dotted vertical line together mark $k_\mt{max}$
        (Equation~\eqref{kmax}).
        Dashed vertical line marks $k_\mt{damp}$, the damping wavelength set by
        PIC current filtering.
    }
    \label{kspectrum}
\end{figure}

\begin{figure}
    \centering
    \includegraphics[width=0.8\linewidth]{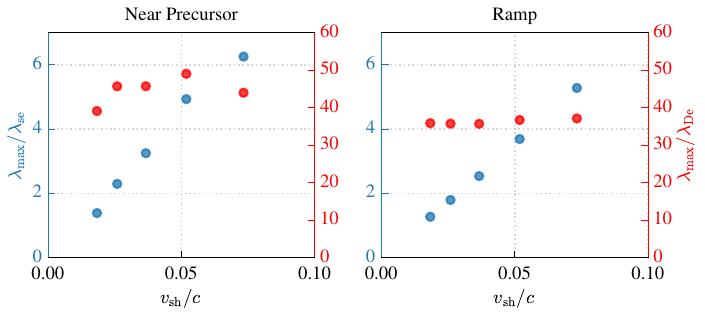}
    \caption{
        The electrostatic power spectrum's peak wavelength
        $\lambda_\mt{max} \equiv 2\pi/k_\mt{max}$
        scaled to $\lse$ (blue dots) and $\lde$ (red dots) for the Near Precursor (left
        column) and Ramp (right column) regions.
    }
    \label{kspectrum_scaling}
\end{figure}

\subsection{Electrostatic wavelengths}

Let us now measure a characteristic electron hole wavenumber
as a function of $v_\mt{sh}/c$,
using the Fourier power spectrum of $E_\parallel/E_0$ for the ``Ramp'' region.
A Hann window function is applied along $x$ and $t$ to reduce
power-spectrum artifacts caused by the signal being aperiodic.
We average the 3D $(k_x, k_y, \omega)$
power spectrum over $\omega$,
and we then sample the 2D power spectrum in $(k_x,k_y)$ by taking a ray along the local $\vec{B}$-field direction ($k_\perp = 0$) within each region.
The resulting 1D spectrum is denoted $W(\vec{k})$ with vector
argument to emphasize that the $k_\perp$ axis is not averaged.

Figure~\ref{kspectrum} shows $W(\vec{k})$ for the ``Ramp'' region of Runs A--E.
The left column of Figure~\ref{kspectrum} scales $k_\prll$ to the electron skin
depth $\lse$;
the right column scales $k_\prll$ to the electron Debye length $\lde$.
The thick, translucent curve is the Fourier power spectrum of the upstream
``Control'' region, which shows the numerical noise floor for comparison.
The vertical dashed line $k_\mt{damp}$ corresponds to a 50\% damping imposed by
the PIC current filtering described in Section~\ref{pic_setup}.
The vertical dotted line shows the peak wavenumber $k_\mt{max}$,
an ensemble-average wavenumber for all the wave power,
defined as:
\begin{equation} \label{kmax}
    k_\mt{max} = \frac{
        \int k_\prll W(\vec{k}=k_\prll \uvec{b}) \,\dtl k_\prll
    }{
        \int W(\vec{k}=k_\prll \uvec{b}) \,\dtl k_\prll
    }
    \, .
\end{equation}
The electrostatic power resides at a fixed multiple of the
electron Debye scale, not the skin depth.
We further affirm this in Figure~\ref{kspectrum_scaling}
by plotting $\lambda_\mt{max}\equiv 2\pi/k_\mt{max}$ as a function of
$v_\mt{sh}/c$, again normalized to either $\lse$ or $\lde$, \artrm{for the ``Ramp'' and ``Near Precursor'' regions.}
By eye, it is clear that $\lambda_\mt{max}/\lse \propto \vsh$, while
$\lambda_\mt{max}/\lde$ does not depend on $\vsh$.
Our measured $\lambda_\mt{max}$ can be reduced by a factor $1/\pi$
to compare to hole lengthscales reported in satellite observations, which
we discuss further in Section~\ref{sec:disc-obs}.

\section{Electron beam model for EAW driving}\label{sec:lin_disp_an}

\rev{Shock simulations from the previous Section demonstrate that electrostatic waves populate both the shock precursor and ramp regions. The amplitude of these waves scales as $(\delta E_\parallel/E_0)^2 \propto (\vsh/c)^{-1}$, while the wavelength scales as $\lambda_\text{max}/\lse \propto \vsh$. In this Section, we clarify the nature and properties of these electrostatic waves using linear dispersion analysis and PIC simulations with periodic boundaries.}

We extract the electron momentum distributions from ``Ramp'' and ``Near Precursor'' regions where electrostatic waves are present.
Electron distributions from each sample region in Run B are illustrated in Figure~\ref{ele_mom_distr}. The distribution in ``Ramp'' and ``Near Precursor'' regions of PIC simulated shocks consists of two hot streams of electrons, a pattern akin to in-situ observations made for Earth's bow shock \citep{Feldman82,Wilson12}.  Notably, the measured in-situ electron distributions can be modeled by a background Lorentzian distribution and a drifting Maxwellian beam. Early theoretical studies \citep{Thomsen83} demonstrated that such electron distribution can drive electron acoustic waves (EAWs) if proper conditions are met. Using simulations of two electron beams with periodic boundaries, we check if distributions extracted from PIC shock simulations are indeed able to drive EAWs, and we compare results with numerical solutions of the hot-beams dispersion relation \artrm{using the code} WHAMP \citep{Ronnmark82}, which employs various approximations of the Fried–Conte plasma dispersion function.

\subsection{Linear dispersion analysis}

\begin{figure}
\centering
\includegraphics[width=\linewidth]{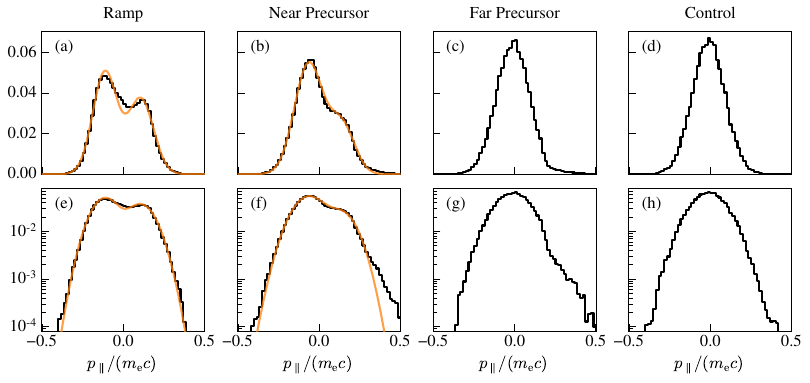}
\caption{
    1D parallel momentum distribution, \aar{integrated over $p_\perp$}, of electrons in the Run B shock regions
    (a) ``Ramp'', (b) ``Near Precursor'', (c) ``Far Precursor'', and (d)
    ``Control''.
    Panels (e)--(h) are the same, but with logarithm-scaled y-axis.
    Black line is average of many \aar{simulation} snapshots ($t=8.00$ to
    $8.50\omci^{-1}$).
    Orange line is best-fit bi-Gaussian distribution (Table~\ref{tab_whamp_param}).
}
\label{ele_mom_distr}
\end{figure}

The distributions in Figure~\ref{ele_mom_distr} can be represented with bi-Gaussian distribution in 1D case. After finding the best fitting Gaussians, we see that the thermal velocities are 2-3 times smaller than the drift velocity ($\vdr/\vth \approx 2-3 $), while the drift velocity is roughly 4 times larger than the shock velocity ($\vdr/\vsh \approx 4 $) \artrm{for simulation with $\mi/\me=200$ (runs A-E)}. Here, the drift velocity $\vdr$ is calculated as the distance between peaks of the two Gaussians and the thermal velocity $\vth$ is the Gaussian's standard deviation, $\vth = \sqrt{k_\mt{B} T_{e,\parallel}/m_\mt{e}}$.
We repeat this fitting procedure for all simulations in Table~\ref{tab_shock_param}.
The best-fit
parameters of electron distributions in Ramp and Near Precursor regions for all simulations are summarised in Table~\ref{tab_whamp_param}. Since the normalized drift and thermal velocities do not depend on the shock velocity, we added a synthetic case (Run S) which \art{is used to extrapolate a realistic shock scenario; it} mimics a run with the average Earth's bow shock velocity of $\vsh = 0.00104c = 312$~km/s \citep{Wilson14}. The parameters of the electron beams \art{($n_1/n_2$, $\vdr/\vsh$, $\vdr/v_{th,1}$, $\vdr/v_{th,2}$)} for Run S were calculated as an average of corresponding values from Runs A--E.

\begin{table}[]
    \begin{tabular}{lccccccccccc}
    \hline
1 & 2 & 3 & 4 & 5& 6& 7& 8& 9& 10\\
    \hline
Run 
    & $\displaystyle \frac{n_1}{n_2}$
    & $\displaystyle \frac{\vdr}{\vsh}$
    & $\displaystyle \frac{\vdr}{v_{th,1}}$
    & $\displaystyle \frac{\vdr}{v_{th,2}}$
    & $\displaystyle \frac{\lambda_{max}}{\lse}$
    & $\displaystyle \frac{\lambda_{max}}{\lde}$
    & $\displaystyle \frac{\Gamma_{max}}{\ompe}$
    & $\left(\displaystyle \frac{\lambda_{max}}{\lse}\right)_{sh}$
    & $\left(\displaystyle \frac{\lambda_{max}}{\lde}\right)_{sh}$  \\
    \hline
\multicolumn{10}{c}{Near Precursor} \\
    \hline
A  & 2.35  & 3.89 & 2.35 & 2.69 & 4.49  & 31.6 & 0.013 & 6.26 & 44.0 \\ 
B  & 2.22  & 3.91 & 2.34 & 2.68 & 3.13  & 31.1 & 0.013 & 4.94 & 49.1 \\ 
C  & 2.25  & 3.87 & 2.33 & 2.92 & 1.82  & 25.6 & 0.029 & 3.25 & 45.8 \\ 
D  & 2.10  & 3.86 & 2.29 & 3.12 & 1.10  & 21.8 & 0.041 & 2.30 & 45.7 \\ 
E  & 1.93  & 3.84 & 2.32 & 3.04 & 0.76  & 21.8 & 0.043 & 1.39 & 39.1 \\ 
S  & 2.17  & 3.87 & 2.33 & 2.89 & 0.052 & 26.0 & 0.025 &  &  \\
    \hline
\multicolumn{10}{c}{Ramp} \\
    \hline
A  & 1.52  & 4.14 & 2.79 & 2.78 & 2.74  & 19.3 & 0.057 & 5.28 & 37.2 \\ 
B  & 1.36  & 4.15 & 2.88 & 2.75 & 1.58  & 15.8 & 0.076 & 3.69 & 36.7 \\ 
C  & 1.33  & 4.16 & 2.96 & 2.61 & 1.36  & 19.2 & 0.060 & 2.54 & 35.7 \\ 
D  & 1.28  & 4.19 & 2.99 & 2.45 & 1.03  & 20.5 & 0.050 & 1.80 & 35.8 \\ 
E  & 1.26  & 4.28 & 3.02 & 2.40 & 0.76  & 21.5 & 0.050 & 1.28 & 35.9 \\ 
S  & 1.35  & 4.19 & 2.93 & 2.59 & 0.04  & 20.1 & 0.049 &   &  \\
B400  & 1.02 & 5.15 & 2.69 & 2.17 & 2.89 & 28.9 & 0.019 &   &  \\
B800  & 0.86 & 6.89 & 2.55 & 1.99 & 10.2 & 102 & 0.00041 &   &  \\
B1836 & 0.66 & 10.2 & 2.42 & 1.76 & -     & -    & stable &   &  \\
    \hline
    \end{tabular}
    \caption{Columns 2--5: parameters of electron beams at “Near Precursor” and "Ramp” regions for all shock simulations.
    Subscript ``1'' corresponds to the denser electron beam moving against magnetic field, and subscript ``2'' corresponds to the diluted electron beam moving along magnetic field.
    Run S is a synthetic run with a realistic shock velocity of $\vsh= 312$km/s \citep{Wilson14}. Columns 6--8: parameters of the most unstable electron acoustic mode according to the linear dispersion analysis.
    Columns 9--10: peak wavelength of the electrostatic power spectrum in shock simulations.}
    \label{tab_whamp_param}
\end{table}

\begin{figure}
\includegraphics[width=.99\linewidth]{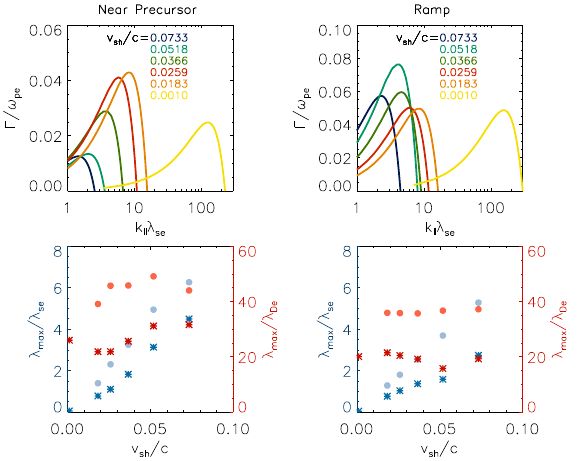}
\caption{Top panels: the growth rate of the EAWs for the near precursor and the shock ramp parameters. Bottom panel: wavelength of the most unstable electron acoustic mode normalised to the electron skin depth (blue asterisks) and the Debye length (red asterisks). Faded red and blue circles show results from shock simulations (also shown in Fig.~\ref{kspectrum_scaling}).}
\label{lin_disp}
\end{figure}

Figure~\ref{lin_disp} shows the growth rate of the EAWs calculated using WHAMP for both regions of interest for all shock simulations and the synthetic case. The growth rate of EAWs falls within the range of $\Gamma_{max}/\ompe \approx 0.01-0.04$ for the ``Near Precursor'' region and  $\Gamma_{max}/\ompe \approx 0.05-0.08$ for the ``Ramp'' region. The frequency of EAWs is in range of $\omega/\ompe=0.3-0.4$ both for ``Near Precursor''  and ``Ramp'' regions. \art{Note that the WHAMP calculations are done in the reference frame of the first beam, where it is stationary, and the second beam moves with $v=\vdr$.}

The growth rate shows slight \rev{variations} across simulations \rev{(compare growth rates for runs A-E)}, although electron beam parameters are very similar. The growth rate is highly sensitive to $\vdr/\vth$ when hot beams are considered. For example, $\Gamma_{max}/\ompe \approx 0.002$ for two beams with $\vdr/\vth = 2.3$, while the growth rate increases by an order of magnitude to $\Gamma_{max}/\ompe \approx 0.028$ when $\vdr/\vth = 2.5$. Nevertheless, shocks generally evolve on ion gyrotime scales, therefore 
\begin{equation} \label{Gamma-omci}
\Gamma/\omci ~\approx (0.01-0.08) \ma  (c/\vsh) \sqrt{\mi/\me} \gg 1
\end{equation} 
indicates that EAWs reach a nonlinear stage in all shock simulations.

Consistent with the findings from shock simulations, the wavelength of the most unstable mode is proportional to the shock speed when normalised to the electron skin depth, $\lambda_{max}/\lse \propto \vsh$, while it remains roughly constant when normalised to the Debye length, $\lambda_{max}/\lde \approx const$. However, the $\lambda_{max}$ values predicted by WHAMP calculations are approximately half of those obtained from shock simulations (see Fig.~\ref{lin_disp}, bottom row). We address this \artrm{discrepancy} in the next subsection.

Parameters of electron beams are influenced by the ion-to-electron mass ratio $\mi/\me$ (runs B, B400, B800, B1836). As the mass ratio increases, \art{the drift velocity also increases relative to $\vsh$ as $\vdr/\vsh \propto \sqrt{\mi/\me}$. However, the average value of the drift velocity relative to the thermal velocity of the drifting electrons decreases when the mass ratio increases}. Consequently, the electron beams become too hot to excite EAWs, as evidenced by the reduced  $\Gamma_{max}/\ompe$ values in Table~\ref{tab_whamp_param}.

\subsection{Periodic boundary condition simulations}

\begin{figure}
\centering
\includegraphics[width=.7\linewidth]{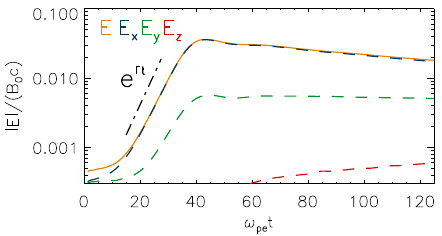}
\caption{Evolution of root-mean-square electric field
fluctuation strength, $|E| = \sqrt{\langle E^2\rangle}$,
in the reference PBCS. The dash-dotted line is the prediction of the linear dispersion analysis, $\Gamma_{max}/\ompe = 0.2$.}
\label{e_evol}
\end{figure}

\begin{figure}
\centering
\includegraphics[width=.7\linewidth]{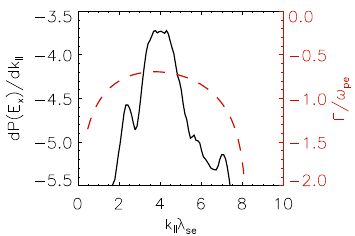}
\caption{The Fourier power spectrum of the electric field parallel to magnetic field in the reference PBCS (black solid line) compared with theoretical prediction (red dashed line).}
\label{e_fourier}
\end{figure}

\begin{figure}
\centering
\includegraphics[width=.7\linewidth]{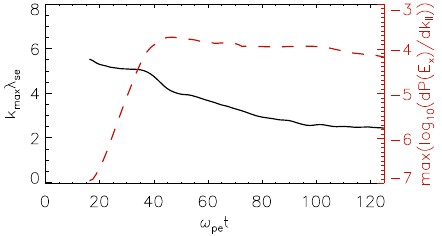}
\caption{Evolution of the
electron acoustic mode $k_\mt{max}$ (black solid line) from its most-unstable wavenumber to a lower-$k$ saturated state, and the peak power (red dashed line) in the reference PBCS.
}
\label{k_evol}
\end{figure}

\begin{figure}
\centering
\includegraphics[width=.99\linewidth]{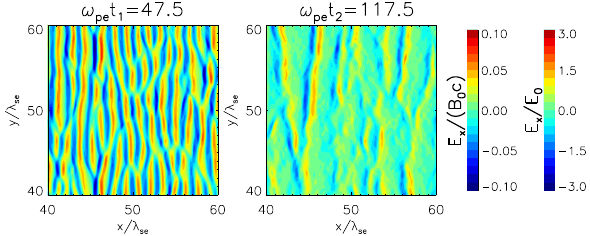}
\caption{The parallel electric field maps at the maximum power of EAWs  (left panel, $\ompe t_1 = 47.5$) and at the late stage of evolution (right panel, $\ompe t_2 = 117.5$) in the reference PBCS.}
\label{epar_map}
\end{figure}

In this section, we explore the evolution of EAWs using 2D periodic-boundary-condition simulations (PBCS). For the initial momentum distribution of electrons, we adopt a bi-Gaussian distribution to represent two hot counterstreaming beams. We initialise two equal density beams with $\vdr/v_{th,1} = 3$ and $\vdr/v_{th,2} = 5$.
These drift speeds exceed the shock-based measurements in Table~\ref{tab_whamp_param} because
we suppose that the electron beams in a shock represent the steady-state
outcome of \artrm{initially unstable conditions}, which may be modeled as having a
larger initial beam drift.
The beams move in opposite directions with magnitudes $|v_1| = |v_2| = \vdr/2$. To comprehensively study the behavior of EAWs, we conduct multiple simulations, varying parameters such as $\vdr$, spatial and temporal resolutions, ion presence/absence and mass ($\mi/\me=200$ and 1836), and the number of particles per cell; \artrm{we keep $\vdr/v_{th,1}$ and $\vdr/v_{th,2}$ constant}.  In the reference run, we use the drift velocity and the strength of magnetic field from Run A. Both the magnetic field and the drift velocities are aligned with the \textit{x}-axis, \rev{mimicking a field-aligned flow that would be inclined with respect to Cartesian coordinate axes in the shock simulations.} Ions are not initialised because they have little if any influence on evolution of EAWs. For the reference run, we set the number of particles per cell per species to $N_{ppc}=2650$ and the spatial grid resolution to $\lse=40\Delta$. Figures \ref{e_evol}, \ref{e_fourier}, \ref{k_evol}, and \ref{epar_map} summarise behaviour of EAWs in the reference PBCS.


Figure~\ref{e_evol} depicts the evolution of the electric field in the reference run, revealing predominantly parallel waves with a minor oblique component. The growth rate is $\Gamma_{max}/\ompe = 0.185$ which closely aligns with the WHAMP prediction of $\Gamma_{max}/\ompe = 0.2$. Slight variations in the growth rate are observed with changes in the number of particles per cell. Increasing $\nppc$ from 40 to 2650 results in a growth rate variation from 0.14 to 0.185. \art{The peak value over time of} the electrostatic field, $max(|E|/(B_0c))$ (where $|E|$ is defined as $\sqrt{\langle E^2 \rangle}$ \art{and the average is taken over the simulation box}), exhibits only marginal changes within the discussed range of $\nppc$, increasing from 0.0341 to 0.0359.
Notably, the growth rate's influence is \art{not} significant, \art{since Equation~\ref{Gamma-omci} is always satisfied} and EAWs have ample time to reach nonlinear stage of evolution.

Figure~\ref{e_fourier} displays the Fourier power spectrum of the electric field parallel to the magnetic field at its maximum intensity ($t\ompe = 47.5$). The peak of the observed spectrum is in good agreement with the numerically-calculated growth rate for the bi-Maxwellian hot beams dispersion relation.
Figure~\ref{k_evol} shows the evolution of $k_\mt{max}\lse$ (see Eq.~\ref{kmax}) in time.
At the time of peak power, the wavelength aligns closely with the WHAMP prediction.
However, during the nonlinear stage of EAW evolution, $k_{\text{max}}\lse$ decreases with time approximately by a factor of two, \rev{explaining the discrepancy in wavelength observed in Figure~\ref{lin_disp}}.
The 2D structure of the EAWs also evolves from
coherent waves at $t_1 \ompe = 47.5$
(Fig.~\ref{epar_map}, left panel)
to bipolar solitary structures at $t_2 \ompe = 117.5$
(Fig.~\ref{epar_map}, right panel).

\begin{figure}
\centering
\includegraphics[width=.7\linewidth]{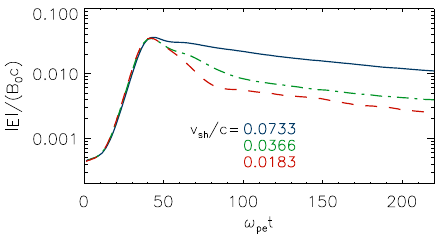}
\caption{The long-term evolution of electric field in the reference PBCS and two runs with $\vdr$ multiplied by 0.5 and 0.25. These PBCS mimic shock conditions from Runs A, C, and E. 
}
\label{kink_infl}
\end{figure}

PBCS demonstrate that the growth rate $\Gamma_{max}/\ompe$ and the maximal electrostatic field strength $max(|E|/(B_0c))$ remain independent of the drift velocity, spatial resolution (if waves are properly resolved, e.g. $\lambda_{max}>10\Delta$), presence or absence of ions, and their mass (assuming $\mi/\me > 200$). However, the long-term evolution reveals that EAWs decay differently depending on the drift velocity. Figure~\ref{kink_infl}  show three PBCS runs with initial conditions drawn from Runs A, C, E (let us call them PBCS A/C/E). \art{These runs demonstrate} different decay behavior at late times \art{($t > 50 \ompe^{-1}$)} where $|E|/(B_0c)$ is roughly proportional to $\vsh/c$ \art{for the chosen time step}.

\section{Discussion} \label{discussion}

The saturated, non-linear outcome of the electron acoustic instability
in PBCS agrees with the full shock simulations in several respects.
The decrease of $k_{\text{max}}\lse$ \artrm{at late times} in Figure~\ref{k_evol} 
can explain why the shock-measured $k_\mt{max}$ differs from the WHAMP
linear predictions by approximately a factor of two
(Table~\ref{tab_whamp_param}, Figure~\ref{lin_disp}).
The polarity of the $E_\prll$ structures in the PBCS
(Figure~\ref{epar_map}) matches that of the shock simulations.
PBCS demonstrate that $\max(|E|/(B_0 c))$ does not depend on $\vsh/c$ \rev{or $(|E| / E_0)^2 \propto (\vsh/c)^{-2}$}. However, at the end of PBCS A/C/E runs, the electrostatic fluctuation amplitude scales as $|E|/(B_0c) \propto \vsh/c$, which implies that $(|E| / E_0)^2 $ is constant with respect to $\vsh/c$.
\rev{Therefore the ESW scaling observed in shock simulations, $\delta E_\prll^2/E_0^2 \propto (\vsh/c)^{-1}$, lies in between the scalings obtained for maximum of electrostatic energy and late-time decay in PBCS.}
Note that $\sqrt{\delta E_\prll^2}$ in shock simulations and $|E|$ in PBCS are almost equivalent, because EAWs in PBCS predominantly generate $E_x$ (see Figure~\ref{e_evol}) which is parallel to the initial magnetic field, therefore $|E| \approx |E_x| \approx \sqrt{\delta E_\prll^2}$. Nevertheless, we continue using $\sqrt{\delta E_\prll^2}$ when referring to shock simulations and $|E|$ when referring to PBCS results.


Let us now further discuss some properties of the late-time
electrostatic wave power in the PBCS and shock simulations,
which we refer to as either ESWs or electron holes. 

\subsection{ESW energy density scaling with $\vsh/c$}

How does the ESW energy density scale with $\vsh/c$ (i.e., $\ompe/\omce$)
for fixed shock parameters (Mach numbers, plasma beta, magnetic obliquity)?
Let the ESW energy density be some fraction $\alpha$ of the electron beams'
drift kinetic energy,
\begin{equation} \label{esw_model}
    \delta E_\prll^2
    \sim \alpha \me n_\mt{e} \vdr^2 / 2
    \, ,
\end{equation}
where $\alpha$ depends on $\vdr/v_{\mt{th},1}$ and $\vdr/v_{\mt{th},2}$.
If the shock's energy partition into various reservoirs---bulk flows, waves, particle heating/acceleration---does not vary with $\vsh/c$ (all other shock parameters held constant), then the ESW energy density, being one of those reservoirs, should scale as $\delta E_\prll^2/E_0^2 \propto (c/\vsh)^2$ \art{and $\max(|E|/(B_0 c))$ should not depend on the shock velocity}. \art{Indeed, we see in shock simulations (runs B-B1836) that} $\vdr^2 \sim (\mi/\me) \vsh^2$, therefore the electrons' drift energy $\me n_\mt{e} \vdr^2/2$ scales linearly with the shock's bulk flow energy $\mi n_\mt{i} \vsh^2/2$ and Equation~\eqref{esw_model} predicts: 

\begin{equation} \label{esw_model_scaling}
    \delta E_\prll^2/E_0^2
    \sim
        \frac{\alpha \mi n_\mt{i} \vsh^2}{(\vsh B_0)^2}
        \sim
        \alpha \beta_\mathrm{p}
        \ms^2
        \left(\frac{c}{\vsh}\right)^2
    \, .
\end{equation}
It suggests that $\delta E_\prll^2/E_0^2$ is independent of the mass ratio $\mi/\me$ for fixed $\alpha$, \rev{which is indeed observed in shock simulations B and B400. Note that $\vsh$ is a factor of 2 different for these simulations, so $\delta E_\prll^2/E_0^2$ is expected to be lower by a factor of 4 in B400.}
In shock simulations \rev{B800 and B1836, however}, the decrease of $\delta E_\prll^2/E_0^2$ with
higher $\mi/\me$ is due to worse driving conditions for ESWs: lower
$\vdr/v_{\mt{th},1}$ and $\vdr/v_{\mt{th},2}$ \rev{(see Table~\ref{tab_whamp_param})} leads to lower $\alpha$, \rev{and numerical noise. It is important to mention that the worsening of driving conditions can potentially be caused by the numerical noise itself.}

The PBCS with initial conditions drawn from Runs A--E agree with
Equation~\eqref{esw_model_scaling}: the same fraction of beam drift kinetic
energy is transferred to \artrm{electrostatic waves} regardless of $\vsh/c$
during the linear-growth stage of the instability,
and when the electric field fluctuation strength
attains its time-series maximum value, $\max(|E|/(B_0 c)) \approx 0.03$ (Figure~\ref{kink_infl}).
\artrm{Why do} the shock simulations and the late-time PBCS runs (Figure~\ref{kink_infl}) show a different scaling?

\begin{figure}
    \centering
    \includegraphics[width=\linewidth]{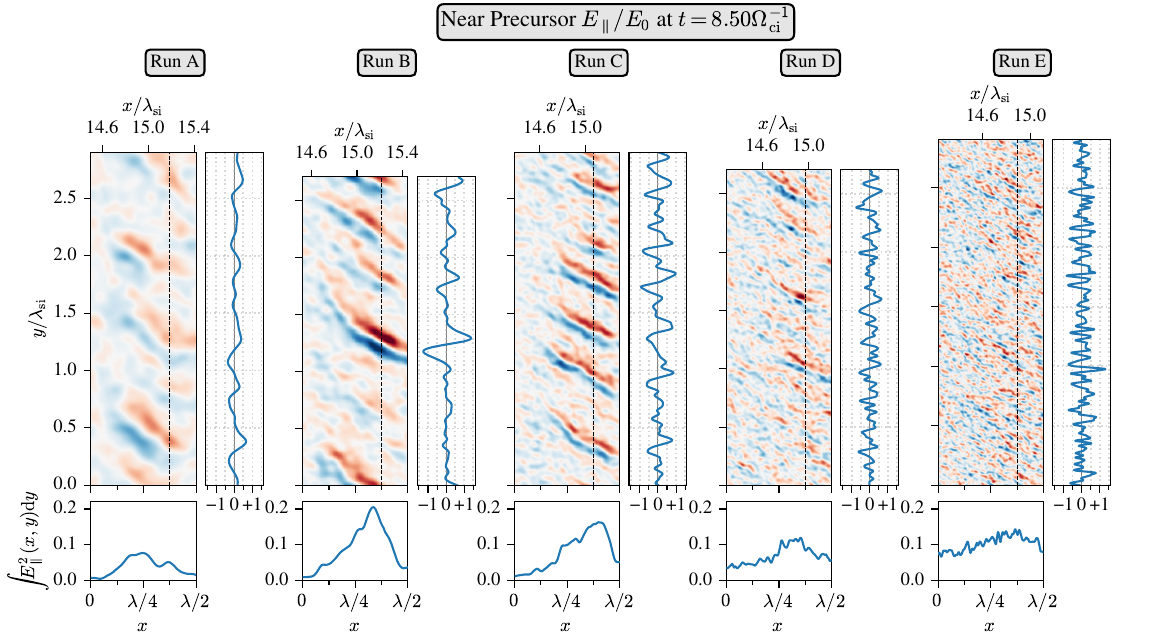}
    \caption{
        Parallel electric field structure in the ``Near Precursor'' region for
        Runs A--E (left to right) at $t=8.50\omci^{-1}$.
        Each column of three panels shows one simulation.
        Within each column, the
        2D image shows $E_\prll/E_0$ with
        the same colormap range as Figure~\ref{ehole}
        and horizontal $x$ axis in units of the
        ion-scale precursor wavelength $\lambda$
        (the precursor's magnetic minimum is at $x=\lambda/4$).
        Right of each 2D image, a plot of $E_\prll(y)$ \artrm{(blue curve)}
        shows the typical amplitude of
        \rev{ESWs} and numerical
        noise at $x=3\lambda/8$ (dashed black line in 2D image).
        Below each 2D image, the mean energy density
        $\int E^2_\prll(x,y) \dtl y$ is plotted as a function of $x$ (blue curve).
    }
    \label{ehole_amplitude}
\end{figure}

First, does the spatial region occupied by the ESWs vary across Runs A--E?
\aar{
In the shock simulations, we measure $\delta E_\parallel^2$ averaged
over an $x$ interval of width $\lambda/2$, where $\lambda$ is the ion-scale precursor wavelength.
If the ESWs occupy an $x$ interval of width $L_x < \lambda/2$,
and $L_x$ varies systematically between Runs A--E,
then the scaling of $\delta E_\parallel^2$ with $\vsh/c$ will be biased
with respect to Equation~\eqref{esw_model_scaling}.
As a concrete example, suppose that
}
$L_x$ is the distance that thermal electrons advect during one EAW instability growth time $\Gamma^{-1}$; i.e.,
$L_x \sim \vsh/\Gamma$.
\aar{Further} suppose that $\Gamma \sim \ompe$ \aar{and that $\Gamma$ is independent of $\vsh/c$.}
Then,
$L_x/\lsi \sim (\ompe/\Gamma) (\me/\mi)^{1/2} (\vsh/c)$ decreases
by a factor of $4$ going from Run A to E,
\aar{so the shock-simulation measurements would be interpreted}
as:

\[
    \left(\frac{\delta E_\prll^2}{E_0^2}\right)_\mt{measured}
    =
    \left(\frac{\delta E_\prll^2}{E_0^2}\right)
        \frac{L_x}{\lambda/2}
    \propto
    \left(\frac{c}{\vsh}\right)^{1}
    \, ,
\]
\aar{
taking $\lambda \approx 2\lsi$ and assuming
$\delta E_\prll^2/E_0^2 \propto (c/v_\mt{sh})^2$
from Equation~\eqref{esw_model_scaling}.
}

But, Figure~\ref{ehole_amplitude} shows qualitatively
that in the ``Near Precursor'' region,
the $y$-averaged parallel electric field power
does not narrow in $x$-width as $\vsh/c$ decreases.
The \rev{electrostatic waves' spatial} width perpendicular to $\vec{B}$
remains constant while the wavelength along $\vec{B}$ shrinks.
Variation in $L_x$ thus does not seem to explain
our measured $\delta E_\parallel^2/E_0^2$ scaling.

Second, does the saturated ESW amplitude vary between Runs A--E, which
would correspond to a change in $\alpha$ in Equation~\eqref{esw_model_scaling}?
Following \citet[Sec.~5]{lotekar2020}
and \citet[Sec.~IV]{kamaletdinov2022},
an electron hole should saturate in amplitude when
\rev{an electron's bounce frequency within the hole's electrostatic potential}
equals either the EAW growth rate $\Gamma$ or the electron cyclotron frequency $\omce$; i.e.,
\begin{equation} \label{eq:hole-saturation}
    \omega_\mt{bounce} = \frac{1}{L} \sqrt{\frac{e\phi}{\me}}
    \le \min(\Gamma, \omce)
    \, ,
\end{equation}
    where $\phi$ is the hole's peak electric potential.
    The case $\omega_\mt{bounce} \sim \Gamma$ is due to non-linear beam
    instability saturation; the prefactor in the scaling relation is somewhat uncertain, as discussed by \cite{lotekar2020}.
    The case $\omega_\mt{bounce} \sim \omce$ is due to hole
    disruption by
transverse instability
\citep{muschietti2000,wu2010,hutchinson2017,hutchinson2018}.
\aar{
These two mechanisms to limit hole amplitudes predict
different scalings of $\delta E_\parallel^2$ with $\vsh/c$.
Taking $\phi \sim \delta E_\parallel L$ and $L \sim \lde$,
Equation~\eqref{eq:hole-saturation} may be rewritten:
}

\[
    e \delta E_\prll \lde \sim \me \Gamma^2 \lde^2
    \, .
\]
If $\omega_\mt{bounce}$ is bounded by  \artrm{$\Gamma \sim \ompe$},
then we find
$\delta E_\prll^2 \sim 4\pi n_\mt{e} k_\mt{B} T_\mt{e}$
which implies a scaling
$\delta E_\prll^2/E_0^2 \propto (\vsh/c)^{-2}$
like Equation~\eqref{esw_model_scaling}.
\aar{
On the other hand, if $\omega_\mt{bounce}$ is bounded by $\omce$, then
$\delta E_\prll^2 \sim 4\pi n_\mt{e} k_\mt{B} T_\mt{e} (\omce/\ompe)^4$
leads to a different scaling
$\delta E_\prll^2/E_0^2 \propto (\vsh/c)^{2}$.
}

\aar{
Taken together, the two mechanisms suggest a non-monotonic scaling
of $\delta E_\prll^2/E_0^2$ with $\vsh/c$.
Recall that lowering $\vsh/c$ towards more realistic values
is equivalent to raising $\ompe/\omce$,
for fixed shock parameters $\ms$ and $\bp$.
For large $\vsh/c$ as in our PIC simulations,
$\Gamma \sim \ompe$ (up to a constant factor) may be less than $\omce$ such that $\delta E_\prll^2/E_0^2$ increases as $\vsh/c$ falls.
Once $\vsh/c$ falls enough so that $\Gamma > \omce$ and transverse instability limits electron hole amplitudes, then
$\delta E_\prll^2/E_0^2$ may peak and then decrease as $\vsh/c$ is further lowered.
}

In our shock simulations,
\aar{
    we estimate
}
$\omega_\mt{bounce} \approx 0.076 \ompe = 0.13 \omce$ (Run A)
and $\omega_\mt{bounce} \approx 0.038 \ompe = 0.27 \omce$ (Run E),
\aar{
    taking $L = 7 \lde$ (Section~\ref{sec:disc-obs})
    and $E_\mt{\prll,peak}/E_0 = 1$
    as typical hole parameters.
}
Both \aar{$\omega_\mt{bounce}$ estimates} lie within the range of $\Gamma_\mt{max} = 0.01$ to $0.08$ for Runs A--E
in Table~\ref{tab_whamp_param},
\aar{
    and both estimates are $\gtrsim 4\times$ smaller than $\omce$.
    The hole amplitudes thus appear to be limited by $\Gamma$ and not transverse instability
    for the range of $\vsh/c$ in our simulations.
}

Electrons in the shock simulations \artrm{are} in steady state.
Could transverse instability have been previously excited
with $\omega_\mt{bounce} > \omce$,
but then stabilized at late times (steady state)
to $\omega_\mt{bounce} < \omce$?
We evaluate this possibility by inspecting PBCS A/C/E
(Figure~\ref{kink_infl}).
At high $\vsh/c$ (PBCS A), the electron holes are long lived, whereas as
$\vsh/c$ decreases the holes disappear.
At $\ompe t=40$, when the electric field energy density is greatest,
the hole amplitude $E_x/(B_0 c) \approx 0.1$ in all PBCS (Figures~\ref{epar_map}, \ref{kink_infl}).
We then estimate $\omega_\mt{bounce} = 0.50 \omce$,
$0.97 \omce$, and $1.96\omce$ for
PBCS A, C, and E respectively, which
suggests that transverse instability may
\artrm{occur during non-linear decay of EAWs into solitary electron holes}
in PBCS C and E.

To summarize, in our shock simulations, \rev{the scaling of the electrostatic energy density associated with the electron holes, 
$\delta E_\parallel^2/E_0^2 \propto (\vsh/c)^{-1}$,} is not well explained by an equipartition argument (Equation~\eqref{esw_model_scaling}).
The hole amplitudes must be \rev{influenced} by non-linear saturation of
electron flows in a manner that is sensitive to $\vsh/c$
(i.e., $\ompe/\omce$).
In matched PBCS simulations, the late-time decay of EAWs into electron holes results in
a scaling of \rev{$|E|/(B_0 c) \propto (\vsh/c)^{-1}$ that corresponds to $\delta E_\prll^2/E_0^2 \propto (\vsh/c)^{0}$, which suggests the importance of non-linear phase for electron holes development in shock simulations.}
The PBCS drives EAW amplitudes large enough that EAW decay into electron holes could be mediated by transverse hole instability.
We speculate that in shocks, an initial electron beam-driving process (e.g., during reflection of a flow off an obstacle) could also form electron holes in such a manner, before settling into the observed steady state.

\subsection{Comparison to observations} \label{sec:disc-obs}

Our shock simulations suggest that the driving conditions for ESWs are independent of the shock velocity (Table~\ref{tab_whamp_param}). Therefore we can expect that these electrostatic waves should be observed in real shocks even if we use $\vsh=312$~km/s as in our synthetic Run S.
The wavelength $\lambda_\mt{max} = 23 \lde \approx 200 \;\mathrm{m}$
for Run S, averaging the values for the Ramp and Near Precursor regions and
assuming $\lde = 8.58 \;\mathrm{m}$ at $1 \unit{AU}$ from the Sun \citep{Wilson21}.
But, we need to make two adjustments.
First, recall that holes in our PBCS runs roughly double in wavelength as the simulation proceeds to late times (Figure~\ref{k_evol}).
Second, our $\lambda_\mt{max} = 2\pi/k_\mt{max}$ does not correspond directly to the hole spatial scale $L$
reported in observations \citep{lotekar2020,kamaletdinov2022}.
The length $L$ arises from a Gaussian model of a hole's electric potential:
\[
    \phi(x) = \phi_0 e^{-x^2/(2L^2)}
    \, .
\]
The Fourier transform of a single hole's $E_\parallel$ signal is
\begin{equation} \label{fourier-hole}
    \tilde{E}_\parallel(k)
    = \frac{1}{\sqrt{2\pi}} \int_{-\infty}^{+\infty} e^{-ikx} E_\prll(x) \dtl x
    = - i \phi_0 L k e^{-k^2 L^2/2}
    \, ,
\end{equation}
and the power spectrum $|\tilde{E}_\prll(k)|^2$ has a local maximum at $k = 1/L$.
All together, we anticipate
$L = 2 \lambda_\mt{max}/(2\pi) \approx 7\lde \approx 60 \;\mathrm{m}$
for our electron holes when scaled to solar wind conditions.
This is comparable to slow electron holes observed at Earth's bow shock---typical
size $\abt 5\lde$, range $\abt 0.5$--$30\lde$ \citep{kamaletdinov2022}---and also
the electron holes seen in Earth's magnetotail \citep{lotekar2020}.

As previously mentioned, PBCS of electron beams show that $max(|E|/(B_0c))$
does not depend on the shock velocity, indicating that $\delta E_\parallel^2 /
E_0^2$ is proportional to $\vsh^{-2}$.
However, in shock simulations, it is observed that
$\delta E_\parallel^2 / E_0^2$ is proportional to $\vsh^{-1}$, or
$\vsh^{0}$ (Figure~\ref{power_scaling}).
By assuming that the true scaling of $\delta E_\parallel^2 / E_0^2$ lies between $\vsh^{-1}$ and $\vsh^{-2}$, and considering that $|\delta E|/E_0$ for individual \rev{ESWs reaches 1.64 in run A}, we can estimate that the amplitude of ESWs in a realistic shock scenario should fall within the range of 
\rev{$|\delta E| \approx 1.64\left(\sqrt{\frac{v_\mathrm{sh,runA}}{v_\mathrm{sh,runS}}}\text{--}\frac{v_\mathrm{sh,runA}}{v_\mathrm{sh,runS}}\right) E_0\approx (14\text{--}116)E_0 \approx (11\text{--}96) mV/m$.} 
These estimates align well with the values measured by MMS \citep{Wilson14b,Goodrich18,Wang21,kamaletdinov2022}. In this estimation, we assumed $B_0=5.8 nT$ \citep{Wilson21}, resulting in $E_0 \approx 0.83 mV/m$.

Our shock simulations suggest that the electron beams become too hot to
drive ESWs as $\mi/\me$ increases towards the true proton-to-electron value
$\mi/\me=1836$.
The beam drift/thermal velocity ratios $\vdr/v_{th,1}$ and $\vdr/v_{th,2}$
decrease monotonically with mass ratio $\mi/\me$ to attain, respectively,
$1.2\times$ and $1.6\times$ smaller values for Run B1836 as compared to Run
B (Table~\ref{tab_whamp_param}).
The instability growth rate $\Gamma_\mt{max}/\ompe$ falls steeply.
But, our simulations at high $\mi/\me$ have strong numerical noise, which reduces distribution anisotropy and hence may bias
our estimates of $\vdr/v_{\mt{th},e}$ low.
And, if the electron beams' drift kinetic energy scales linearly with
the shock frame's incoming bulk energy $\vsh^2$, implying
$\vdr/v_{\mt{th},e} \propto \ms$, a $1$-$2\times$ increase in $\ms$
may suffice to drive EAW-unstable beam drifts in a shock with realistic
mass ratio.

In both shock simulations and PBCS, we observe electron holes with positive polarity (net positive charge and local electric potential maximum).
For our chosen shock parameters, few ions reflect at the ramp and the
overall shock structure is laminar, so ion-ion streaming does not occur and
ion holes of negative polarity (net negative charge and local electric potential minimum)
are not generated.
In contrast, the bipolar ESWs observed at Earth's bow shock are mostly ion
holes.
\citep{Wang21} present a detailed catalog of bipolar ESWs measured in ten
MMS crossings of Earth's bow shock; in eight crossings, electron holes are
only $1$-$6\%$ of the catalogued bipolar ESWs.
However, in the two crossings with lowest $\ma=3.4$ and $4.7$, electron
holes are $\abt 25\%$ of the catalogued bipolar ESWs.
Our simulations are thus most pertinent to lower-Mach crossings of Earth's
bow shock and interplanetary shocks in the heliosphere.
Further shock simulations encompassing different Mach numbers \artrm{and obliquities} are necessary to investigate the nature of the various ESWs observed near Earth's bow shock region.


\section{Conclusions}

In this study, we have identified ESWs observed in low Mach number shock simulations as the non-linear outcome of the electron acoustic instability, which has been confirmed through simulations with periodic boundaries and linear dispersion analysis. These ESWs are driven by two hot counter-streaming electron beams, and \art{the ratio of the drift velocity to the thermal velocity for these beams (in other words, driving conditions)} is independent of the shock velocity. This finding suggests that the same mechanism can be responsible for driving ESWs in shocks with realistic velocities. Additionally, we have observed that the wavelength of ESWs is proportional to the shock velocity, and this expected wavelength under Earth's bow shock conditions is consistent with in-situ measurements obtained by MMS. Furthermore, we have found that the normalized strength of ESWs is roughly inversely proportional to the shock velocity, indicating that in real shocks, their amplitudes would be significantly higher than the quasi-static electric field, aligning with observations from in-situ measurements.
However, the usage of the realistic proton-to-electron mass ratio alters the driving conditions and strongly suppresses the occurrence of ESWs \art{for the Mach number and shock obliquity chosen in our study}. This suppression is due to a combination of the high electron thermal velocity in comparison to the drift velocity of the two electron beams and \art{significant} numerical noise.
Better-quality shock simulations are needed to accurately measure the
drift/thermal velocity ratio and to suppress numerical noise, in order to
assess whether \art{this particular type of} ESW may be driven at the true mass ratio.
And, in stronger shocks \rev{(higher $\ms$)} at $\mi/\me=1836$, we also anticipate that
higher drift velocities and local (shock-transverse) fluctuations in
electron beam driving may drive ESWs.

This study proposes a solution for the discrepancy between PIC simulations and in-situ measurements. In PIC simulations, we \rev{observe} ESWs with parameters (wavelength and amplitude) \rev{that differ from those observed at the Earth's bow shock due to the higher shock velocity used in PIC simulations}. However, the nature of the ESWs can be the same as in real shocks. These conclusions can be applied to similar two-stream electrostatic instabilities, \rev{such as IAWs, if shocks with different Mach numbers or obliquity are considered}. While our focus has been on beams induced by large-amplitude oblique whistlers, this sort of beam-beam interaction and electrostatic wave generation is a generic process. If the driving conditions remain constant across shocks with varying velocities and ion-to-electron mass ratios, the electrostatic waves observed in PIC simulations may appear in real shocks with the correct strength and wavelength.
\rev{Additionally, it is important to highlight that the small amplitude and large wavelength electrostatic waves observed in PIC simulations are a realistic representation of electrostatic waves for the chosen shock velocity, provided that numerical noise is negligible.}

\art{\section*{Data Availability Statement}}

\art{Public versions of Tristan and WHAMP are available at
\url{https://github.com/ntoles/tristan-mp-pitp}
and \url{https://github.com/irfu/whamp}.
The scripts required to generate the model data and the figures in this paper are available at https://doi.org/10.5281/zenodo.10973653 \cite{bohdan_2024}.}

\begin{acknowledgments}
A.B. was supported by the German Research Foundation (DFG) as part of the Excellence Strategy of the federal and state governments - EXC 2094 - 390783311. A.T. was supported by NASA FINESST 80NSSC21K1383
\aar{and partly supported by NSF PHY-2010189.}
\aar{L.S. was supported by NASA ATP grant 80NSSC20K0565.}
Some of the work was supported by the Geospace Environment Modeling Focus Group ``Particle Heating and Thermalization in Collisionless Shocks in the Magnetospheric multiscale mission (MMS) Era'' led by L.B. Wilson III. We acknowledge travel and collaboration support from the ISSI Bern Team ``Energy Partition across Collisionless Shocks'' (ISSI International Team project \#520). We thank Martin Pohl and Ivan Vasko for helpful discussion. The numerical experiments were done with HLRN supercomputer at North-German Supercomputing Alliance (projects bbp00033 and bbp00057), the HPC system Raven at the Max Planck Computing and Data Facility (MPCDF), the computer clusters Habanero, Terremoto, Ginsburg (Columbia University), and Pleiades allocation s2610 (NASA). The latter resources were provided by Columbia University's Shared Research  Computing Facility (SRCF) and the NASA High-End Computing Program through the NASA Advanced Supercomputing (NAS) Division at Ames Research Center. Columbia University's SRCF is supported by NIH Research Facility Improvement Grant 1G20RR030893-01 and the New York State Empire State Development, Division of Science Technology and Innovation (NYSTAR) Contract C090171.
\end{acknowledgments}

\appendix

\section{Precursor $E_\parallel$ power spectrum in Runs A--E} \label{app1}

\begin{figure}
\centering
\includegraphics[width=0.9\linewidth]{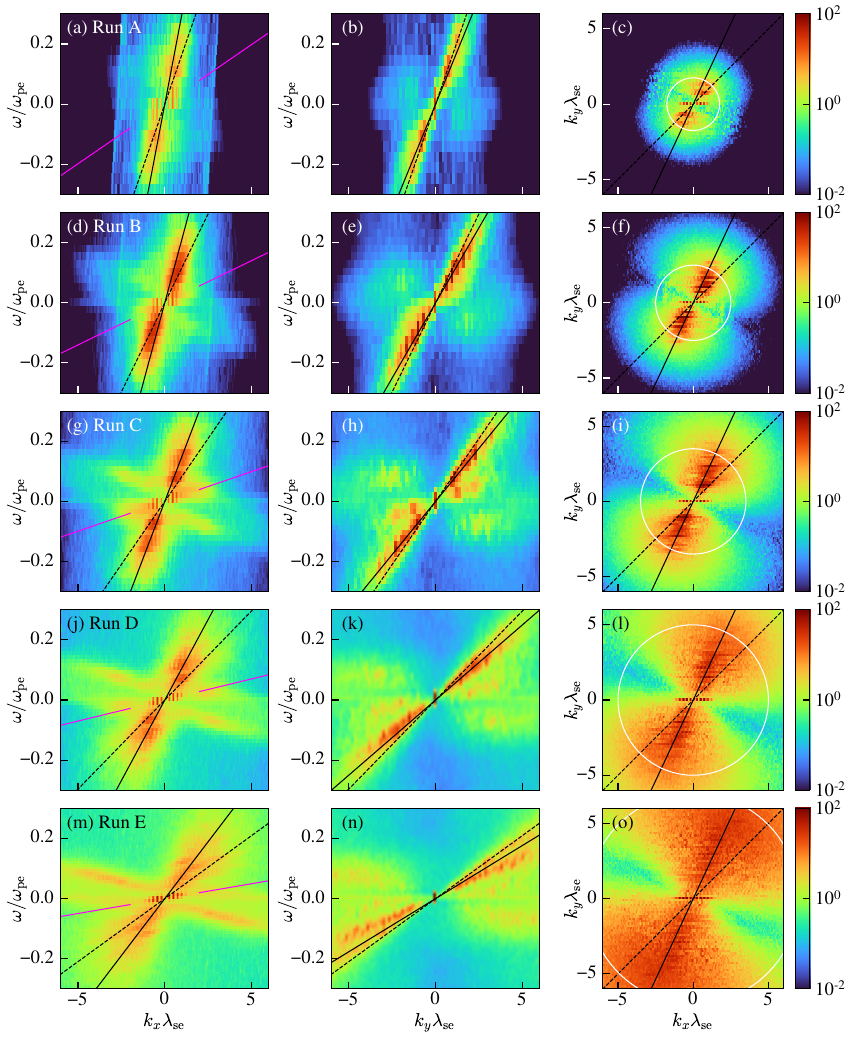}
\caption{
    Fourier power spectra of shock-precursor $E_\parallel$
    for Runs A--E (top to bottom),
    \aar{measured in the downstream plasma's rest frame.}
    Left column is ($\omega$, $k_x$) spectrum,
    middle column is ($\omega$, $k_y$) spectrum,
    and right column is ($k_x$, $k_y$) spectrum.
    \aar{Each column is averaged over all $k_y$, $k_x$, and $\omega$ respectively.}
    The low-frequency,\aar{phase-standing}
    precursor
    lies along $\omega/k_x = v_\mt{sh}/r$ (left column, magenta line)
    where $r$ is the shock's \aar{density} compression ratio;
    \aar{near the origin, the magenta line is not drawn so that
    features of interest can be seen.}
    A broad region of wave power
    \aar{
        has $\vec{k}$ along $\vec{B}$
        and downstream-frame phase velocity $\omega/k$ that,
        when boosted to the upstream rest frame, is close to the upstream electron thermal speed $v_\mt{te0}$.
        To show this,
    }
    solid and dashed black lines plot
    $\omega = k v_\mt{te0} - k_x v_0$ at propagation angles
    $\theta = 65^\circ$ and $45^\circ$ respectively, measured counterclockwise from $+\uvec{x}$
    \aar{
        (i.e.,
        $\omega = k_x v_\mt{te0}/\cos\theta - k_x v_0$
        (left column) and
        $\omega = k_y v_\mt{te0}/\sin\theta - k_y v_0/\tan\theta$
        (middle column)).
    }
    The $\theta$ angles bracket the range of $\vec{B}$
    \aar{orientations} within precursor wave troughs.
    White circle (right column) is 50\% damping length induced by PIC current
    filtering (Section~\ref{pic_setup}).
    }
\label{shock_fft_eprll}
\end{figure}

The Fourier power spectrum of $E_\parallel$ also confirms the propagation
direction and speed of the electron holes (Figure~\ref{shock_fft_eprll}),
\aar{
    serving a similar purpose as Figure~\ref{ehole} but without needing to manually track individual holes.
}
We compute the spectrum for all of Runs A--E
in the region $x=10$ to $20\lsi$ and time interval $t=8.00$ to $t=8.50\omci^{-1}$;
\aar{
the measurement is performed in the downstream rest frame (i.e., simulation frame).
}
The time sampling rate $f \approx 0.444 \ompe$ resolves Langmuir wave power
at $\omega \approx \ompe - k_x v_0$ in the simulation frame,
ensuring that it does not alias in frequency space
and thereby contaminate the $E_\prll$ spectral power of interest to us.
The Langmuir waves' Doppler shift $k_x v_0$ is $\lesssim 20\%$ of $\ompe$ at $k = 5 \lse^{-1}$ for all runs, so the waves are unaliased and well separated
for the $k$ domain in Figure~\ref{shock_fft_eprll}.
In all of Runs A--E, we observe $E_\parallel$ wave power at $\omega = 0$ to
$0.2\ompe$ that clusters along
$\omega = k_\parallel v_\mt{te0}$.
The waves occupy a broad bandwidth in both $\omega$ and $k$, which at high $k$
is limited by the damping lengthscale of the PIC current filtering.

\bibliography{agusample,library_aaron}

\bibliographystyle{aasjournal}

\end{document}